\documentclass[12pt]{iopart}
\usepackage{epsfig}
\usepackage{citesort}

\newcommand{\gtrsim}{\,\rlap{\lower3.7pt\hbox{$\mathchar\sim$}}
\raise1pt\hbox{$>$}\,}
\newcommand{\lesssim}{\,\rlap{\lower3.7pt\hbox{$\mathchar\sim$}}
\raise1pt\hbox{$<$}\,}

\begin{document}

\title{Cosmological constraints on the dark energy equation of state
and its evolution}
\author{Steen Hannestad \footnote{hannestad@fysik.sdu.dk}}
\address{Physics Department, University of Southern Denmark\\
Campusvej 55, DK-5230 Odense M, Denmark \\
and \\
Theory Division, CERN, CH-1211 Geneva 23, Switzerland}
\author{Edvard M{\"o}rtsell \footnote{edvard@astro.su.se}}
\address{Department of Astronomy, Stockholm University, SE-106 91 Stockholm, Sweden}

\date{{\today}}

\begin{abstract}
We have calculated constraints on the evolution of the equation of state of the dark energy, $w(z)$, from a joint analysis of data from the cosmic microwave background, large scale structure and type-Ia supernovae.
In order to probe the time-evolution of $w$ we propose a new, simple parametrization of $w$, which has the advantage of being transparent and simple to extend to more parameters as better data becomes available. Furthermore it is well behaved in all asymptotic limits. Based on this parametrization we find that $w(z=0)=-1.43^{+0.16}_{-0.38}$ and $dw/dz(z=0) = 1.0^{+1.0}_{-0.8}$. For a constant $w$ we find that $-1.34 \leq w \leq -0.79$ at 95\% C.L. Thus, allowing for a time-varying $w$ shifts the best fit present day value of $w$ down. However, even though models with time variation in $w$ yield a lower $\chi^2$ than pure $\Lambda$CDM models, they do not have a better goodness-of-fit. Rank correlation tests on supernova data also do not show any need for a time-varying $w$.
\end{abstract}
\maketitle

\section{Introduction} 

The discovery in 1998 \cite{Riess:1998cb,Perlmutter:1998np} that the universal expansion is currently accelerating is one of the most spectacular result in cosmology from the past decade. The finding has since been confirmed by observations of the Cosmic Microwave Background (CMB) \cite{Bennett:2003bz,Spergel:2003cb} and the large scale structure (LSS) of the universe \cite{2dFGRS,Tegmark:2003uf,Tegmark:2003ud}

One possible explanation is that the energy density of the universe is dominated by dark energy with a negative equation of state.
The simplest possibility is the cosmological constant which has $P=w\rho$ with $w=-1$ at all times. However, since the cosmological constant has a value completely different from theoretical expectations one is naturally led to consider other explanations for the dark energy.

A light scalar field rolling in a very flat potential would for instance have a strongly negative equation of state, and would in the limit of a completely flat potential lead to $w=-1$ \cite{Wetterich:1987fm,Peebles:1987ek,Ratra:1987rm}. Such models are generically known as quintessence models. The scalar field is usually assumed to be minimally coupled to matter, but very interesting effects can occur if this assumption is relaxed (see for instance \cite{Mota:2004pa}).

In general such models would also require fine tuning in order to achieve $\Omega_X \sim \Omega_m$, where $\Omega_X$ and $\Omega_m$ are the dark energy and matter densities at present.
However, by coupling quintessence to matter and radiation it is possible to achieve a tracking behavior of the scalar field so that $\Omega_X \sim \Omega_m$ comes out naturally of the evolution equation for the scalar field \cite{Zlatev:1998tr,Wang:1999fa,Steinhardt:1999nw,Perrotta:1999am,%
Amendola:1999er,Barreiro:1999zs,Bertolami:1999dp,Baccigalupi:2001aa,%
Caldwell:2003vp}.

Many other possibilities have been considered, like $k$-essence, which is essentially a scalar field with a non-standard kinetic term \cite{Armendariz-Picon:1999rj,Chiba:1999ka,Armendariz-Picon:2000ah,%
Chimento:2003ta,Gonzalez-Diaz:2003rf,Scherrer:2004au,Aguirregabiria:2004te}.
It is also possible, although not without problems, to construct models which have $w<-1$, the so-called phantom energy models 
\cite{Caldwell:1999ew,Schulz:2001yx,Carroll:2003st,Gibbons:2003yj,%
Caldwell:2003vq,Nojiri:2003vn,Singh:2003vx,Dabrowski:2003jm,Hao:2003th,%
Stefancic:2003rc,Cline:2003gs,Brown:2004cs,Onemli:2002hr,Onemli:2004mb,%
Vikman:2004dc}.

Finally, there are even more exotic models where the cosmological acceleration is not provided by dark energy, but rather by a modification of the Friedman equation due to modifications of gravity on large scales \cite{Deffayet:2001pu,Dvali:2003rk}.

Given this plethora of different possibilities and since we have no fundamental theory available for calculating it appears that $w$ should be treated as an effective parameter only. In many models $w$ is also changing with time, typically going from one asymptotic limit for $z \to \infty$ to another for $1+z \to 0$.

The simplest parametrization is $w=$ constant, for which constraints based on observational data have been calculated many times
\cite{Corasaniti:2001mf,Bean:2001xy,Hannestad:2002ur,Melchiorri:2002ux}. However, as the precision of observational data is increasing is it becoming feasible to search for time variation in $w$.

In the present paper we propose a very simple parametrization for $w$ which allows us to treat almost all models for dark energy currently on the market. Furthermore it is straightforward to extend our parametrization when better data becomes available, and the parametrization is well-behaved in all relevant asymptotic limits.
The paper is structured as follows: In section 2 we describe our parametrization and its relation to supernova luminosity distances and CMB angular distances. In section 3 we discuss how a numerical likelihood analysis has been performed with the most recent cosmological data, and in section 4 we discuss our results. We also discuss how to extend our current parametrization to provide a more refined description of $w(z)$ when more data becomes available. Finally, section 5 contains a conclusion.

\section{Luminosity distance, angular distance, and the parametrization of $w$} 

\subsection{Supernova luminosity distances}
Type Ia supernova (SNIa) observations provide the currently most
direct way to probe the dark energy at low to medium redshifts since
the luminosity-distance relation is directly related to the expansion
history of the universe.

The luminosity distance $d_L$ is given by
\begin{eqnarray}
    d_L&=&\left\{
    \begin{array}{ll}
      (1+z)\frac{1}{\sqrt{-\Omega_k}}\sin(\sqrt{-\Omega_k}\,I) , &
      \Omega_k<0\\
      (1+z)\,I , & \Omega_k=0\\
      (1+z)\frac{1}{\sqrt{\Omega_k}}\sinh(\sqrt{\Omega_k}\,I) , &
      \Omega_k>0\\
    \end{array}
    \right. \label{eq:snl} \\
    \Omega_k&=&1-\Omega_m-\Omega_X,\nonumber \\
    I&=&\int_0^z\,\frac{dz'}{H(z')} ,\nonumber \\
    H(z)&=H_0&\sqrt{(1+z)^3\,\Omega_{m}+
    f(z)\,\Omega_{X}+(1+z)^2\,\Omega_{k}}, \nonumber \\
    f(z)&=&\exp\left[3\int_0^z\,dz'\,\frac{1+w(z')}{1+z'}\right] \nonumber .
\end{eqnarray}
Here, $w(z)$ is an arbitrary function of redshift. Putting $w={\rm
constant}=-1$, we have $\Omega_X=\Omega_\Lambda$ and the luminosity
distance relation can be used to constrain the values of $\Omega_m$ and
$\Omega_\Lambda$. Using current SIa data,
$\Omega_\Lambda=0$ can be ruled out at very high confidence level
\cite{Knop:2003,Riess:2004} (note that one marginalizes over an arbitrary
multiplicative factor in $d_L$ and that the Hubble parameter $H_0$ is
thus treated as a free parameter).

In principle, it should also be possible to constrain $w(z)$ using
SNIa data. In practice however, this is quite complicated since the
luminosity distance depends on $w$ through multiple integrals causing
degeneracies between parameters \cite{Maor:2001}. Also, the sheer number of
parameters to fit in relation to the quantity and quality of current
data makes this difficult to do. As a simple rule of thumb, it is
hard to constrain more than two parameters using current SNIa
data. A common approach is to assume a flat universe, i.e.,
$\Omega_X=1-\Omega_m$ and fit $\Omega_m$ and $w={\rm constant}$. The
flat universe assumption can be justified from a theoretical point of
view as a consequence of inflation with observational support from CMB
angular scale measurements. This approach gives an limit of $w\lesssim
-0.5$ using the ``gold'' sample of 157 SNIa compiled in
\cite{Riess:2004}. Combining the SN data with CMB and LSS measurements
yields $w=-1.08\pm^{0.20}_{0.18}$, consistent with a cosmological
constant.

The next step is to constrain the time variation in the equation of
state parameter. One approach is to express $w$ as a first order
Taylor expansion in $z$ (usually around $z=0$)\footnote{Of course, $w$
can be expanded to higher order with the price of adding more
parameters to the fit.}
\begin{equation}
w(z)=w|_{z=0}+\frac{dw}{dz}\Big |_{z=0}\,z .
\end{equation}
The advantage of this method is the simplicity; any value of
$dw/dz(z=0) \neq 0$ would indicate a non-constant
equation of state parameter. The obvious disadvantage is the
divergence of $w$ at high redshifts that makes it practically useful
for SN data only\footnote{A similar parametrization that avoids this
behaviour at high redshift is $w(a)=w|_{a=1}+w_a(1-a)$
\cite{Linder:2003}. However, that parametrization diverges at $a \to \infty$.}. Assuming a flat universe and a prior of
$\Omega_m=0.27\pm 0.04$ (derived assuming a cosmological constant),
Riess et al. obtains $w|_{z=0}=-1.31\pm^{0.22}_{0.28}$ and
$dw/dz(z=0)=1.48\pm^{0.81}_{0.90}$ with the case of a
cosmological constant within the (joint) 68\,\% confidence level.

In additional to the constant and linear models of $w$ there exists an
ever increasing number of models for $w$ that can be broadly
classified into parametric models
\cite{Corasaniti:2002,Corasaniti:2004,Gong:2004a,Gong:2004b,Nesseris:2004wj,%
Feng:2004ad,Alam:2004a,Alam:2004b,Jassal:2004ej} and non-parametric models
\cite{Huterer:2004,Daly:2003,Wang:2004a,Wang:2004b,Wang:2004c}.

There are large differences in how to include information from CMB and
LSS data, the choice of SN data sets, the marginalization over the SN
magnitude zero point, the use of e<xternal priors in $\Omega_m$ and
$H_0$ etc. In spite of this, the consensus is that current data are
consistent with dark energy in the form of a cosmological constant at
the $2\sigma$ level. The best fit model for a time varying equation of
state tend to have lower values of $w$ at low redshift. Whether this
is a real effect or an artefact of the specific model employed in the
fit is currently a subject of discussion
\cite{Jonsson:2004,Alam:2004b}, but the fact remains that any indications of a possible
time evolution in $w$ are still statistically insignificant.

In order to be able to constrain the temporal variation of $w$, it is
crucial to combine different observational data in a self consistent
way. For example, using priors from CMB and LSS derived assuming a
cosmological constant when constraining $w$ using SN data may lead
inconsistencies. However, constraining $w$ over a redshift range of
$0<z\lesssim 1000$ puts high demands on the parametrization (or
non-parametric model) of $w$.

\subsection{Parametrization of $w(a)$}

In order to describe a possible time evolution of $w(z)$, we propose
the following very simple approximation

\begin{equation}\label{eq:wa}
w(a) = w_0 w_1 \frac{a^q + a_s^q}{a^q w_1 + a_s^q w_0},
\end{equation}
where $a$ is the scale factor normalized to the value of one today,
i.e, $a=(1+z)^{-1}$ and $w_0,w_1,a_s$, and $q$ are constants.  $w_0$
and $w_1$ describe the asymptotic behaviour of $w$
\begin{equation}
w(z) \to \cases{w_0 & for $1+z \to 0$ \cr
w_1 & for $1+z \to \infty$}.
\end{equation}

\begin{figure}
\hspace*{1.5cm}\includegraphics[width=100mm]{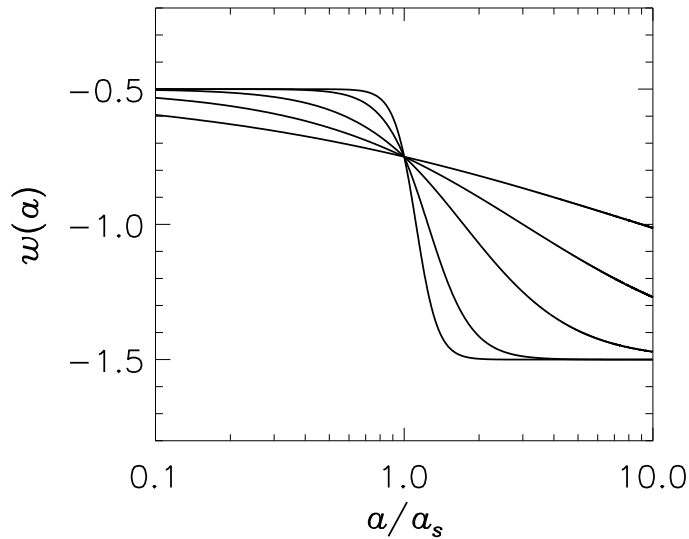}
\caption{$w(a/a_s)$ for different values of $q$. The values shown are $q = 0.5, 1,2,5,10$ in order of increasing steepness.}
\label{fig:q}
\end{figure}

The two additional parameters $a_s$ and $q$ describe the scale factor
at changeover and the duration of the changeover in $w$
respectively. In figure \ref{fig:q} we show the evolution of $w$ as a
function of redshift for different valus of $q$ ranging from 0.5 to 10,
the lower the value of $q$, the longer the duration of the changeover
in $w$.  In this plot, we have fixed $w_0=-0.5$ and $w_1=-1.5$.

Advantages of this parametrization include the simple and intuitive
role of each parameter as well as the large flexibility that aids in
avoiding problems such as ``sweet spots'' and fixed asymptotic values
for $w$.

\subsection{CMB angular scales}

Changing the equation of state of the dark energy while keeping all
other parameters of the model fixed has two distinct effects: a) It
changes the magnitude of the ISW effect at low $l$, b) It introduces a
linear shift in the angular position of the CMB features by a factor
$F$, given by \cite{Melchiorri:2002ux}
\begin{equation}
F = \Omega_m^{1/2} \int_0^{z_{rec}} [\Omega_m (1+z)^3 + (1-\Omega_m)
(1+z)^{3(1+w(z))}]^{-1/2}.
\end{equation}

\begin{figure}
\hspace*{1.5cm}\includegraphics[width=100mm]{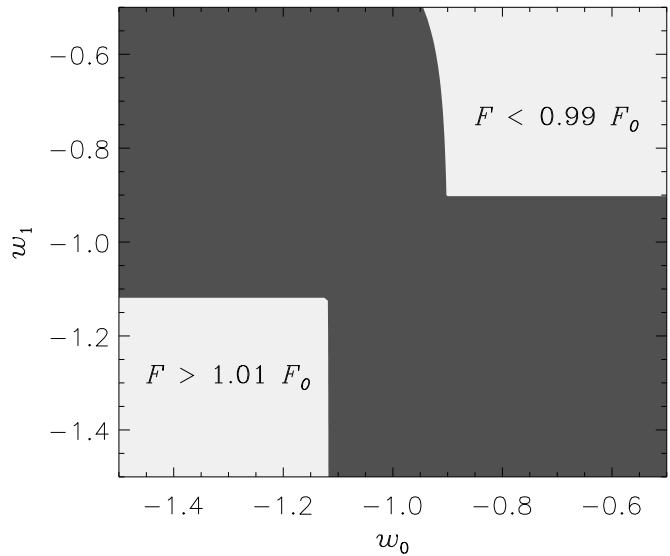}
\caption{Contourplot of the function $F$. For each value of 
$(w_0,w_1)$, $a_s$ and $q$ have been chosen so that $F$ is as close to
$F_0 = 1.61$ as possible.}
\label{fig:angle}
\end{figure}

In figure \ref{fig:angle} we show the value of $F$ relative to that of
an $\Omega_m = 0.3$ flat $\Lambda$CDM model. For each value of $(w_0,w_1)$, $a_s$ and $q$ have been
chosen so that $F$ is as close to $F_0 = 1.61$ as possible.  From the
figure it can be seen that a significant degeneracy in $(w_0,w_1)$ can
be expected for CMB data. The reason simply is that $F$ is an integral
along a line of sight so that changing $w$ by different amounts at
different epochs can yield the same integrated result. CMB does not
have access to the instantaneous value of $w$ at each given redshift
point, only to the integral $F$.

\section{Numerical analysis}

Since it is not possible to obtain useful limits for all four
parameters of our equation of state parametrization in equation
\ref{eq:wa} and we are primarily interested in exploring 
evidence for $w\neq -1$ and/or a changing $w$, we employ the following
approach: For fixed values of $w_0$ and $w_1$ we marginalize over
the other cosmological parameters. As our framework we choose the
minimum standard model with 6
parameters: $\Omega_m$, the matter density, the curvature
parameter, $\Omega_b$, the baryon density, $H_0$, the Hubble
parameter, and $\tau$, the optical depth to reionization. The normalization of both CMB and LSS spectra are taken to be free and unrelated parameters.
The priors we use are given in Table~\ref{tab:priors}.

\begin{table}[ht]
\caption{\label{tab:priors} Priors on cosmological parameters used in
the likelihood analysis.}
\begin{indented}
\item[]
\begin{tabular}{@{}lll}
\br
Parameter &Prior&Distribution\cr
\mr
$\Omega=\Omega_m+\Omega_X$&1&Fixed\\
$h$ & $0.72 \pm 0.08$&Gaussian \cite{freedman}\\
$\Omega_b h^2$ & 0.014--0.040&Top hat\\
$n_s$ & 0.6--1.4& Top hat\\
$\tau$ & 0--1 &Top hat\\
$Q$ & --- &Free\\
$b$ & --- &Free\\
\br
\end{tabular}
\end{indented}
\end{table}

Likelihoods are calculated from $\chi^2$ so that for 1 parameter estimates, 68\% confidence regions are determined by $\Delta \chi^2 = \chi^2 - \chi_0^2 = 1$, and 95\% region by $\Delta \chi^2 = 4$. $\chi_0^2$ is $\chi^2$ for the best fit model found.
In 2-dimensional plots the
68\% and 95\% regions are formally
defined by $\Delta \chi^2 = 2.30$ and 6.17
respectively. Note that this means that the 68\% and 95\% contours
are not necessarily equivalent to the same confidence level for single
parameter estimates.

\subsection{Supernova luminosity distances}
We perform our likelihood analysis using the ``gold'' dataset compiled
and described in Riess et al \cite{Riess:2004} consisting of 157
SNIae\footnote{Note that the electronic table in ApJ is one SN
short. To get the full data set, use the table in astro-ph
version. The missing SN is SN1991ag.} using a modified version of the
SNOC package \cite{Goobar:2002c}.

\subsection{Large Scale Structure (LSS).}

At present there are two large galaxy surveys of comparable size, the
Sloan Digital Sky Survey (SDSS) \cite{Tegmark:2003uf,Tegmark:2003ud}
and the 2dFGRS (2~degree Field Galaxy Redshift Survey) \cite{2dFGRS}.
Once the SDSS is completed in 2005 it will be significantly larger and
more accurate than the 2dFGRS. In the present analysis we use data from SDSS, but the results would be almost identical had we used 2dF data instead. In the data analysis we use only data points on scales larger than $k = 0.15 h$/Mpc in order to avoid problems with non-linearity.

\subsection{Cosmic Microwave Background.}

The CMB temperature fluctuations are conveniently described in terms of
the spherical harmonics power spectrum $C_l^{TT} \equiv \langle
|a_{lm}|^2 \rangle$, where $\frac{\Delta T}{T} (\theta,\phi) =
\sum_{lm} a_{lm}Y_{lm}(\theta,\phi)$.  Since Thomson scattering
polarizes light, there are also power spectra coming from the
polarization. The polarization can be divided into a curl-free $(E)$
and a curl $(B)$ component, yielding four independent power spectra:
$C_l^{TT}$, $C_l^{EE}$, $C_l^{BB}$, and the $T$-$E$ cross-correlation
$C_l^{TE}$.

The WMAP experiment has reported data only on $C_l^{TT}$ and $C_l^{TE}$
as described in
Refs.~\cite{Bennett:2003bz,Spergel:2003cb,%
Verde:2003ey,Kogut:2003et,Hinshaw:2003ex}.  We have performed our
likelihood analysis using the prescription given by the WMAP
collaboration~\cite{Spergel:2003cb,%
Verde:2003ey,Kogut:2003et,Hinshaw:2003ex} which includes the
correlation between different $C_l$'s. Foreground contamination has
already been subtracted from their published data.

\section{Results}

\begin{figure}
\hspace*{1.5cm}\includegraphics[width=100mm]{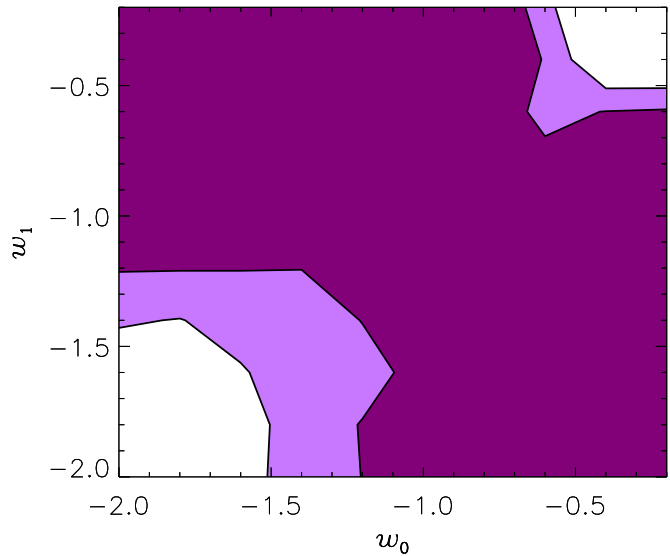}
\caption{68\% (dark) and 95\% (light) likelihood contours for WMAP and SDSS data only.}
\label{fig:chi1}
\end{figure}

\begin{figure}
\hspace*{1.5cm}\includegraphics[width=100mm]{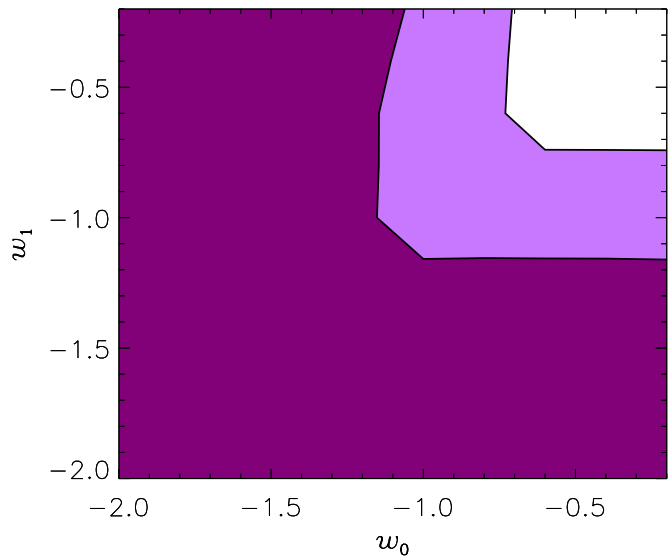}
\caption{68\% (dark) and 95\% (light) likelihood contours for SNI-a data only.}
\label{fig:chi3}
\end{figure}

\begin{figure}
\hspace*{1.5cm}\includegraphics[width=100mm]{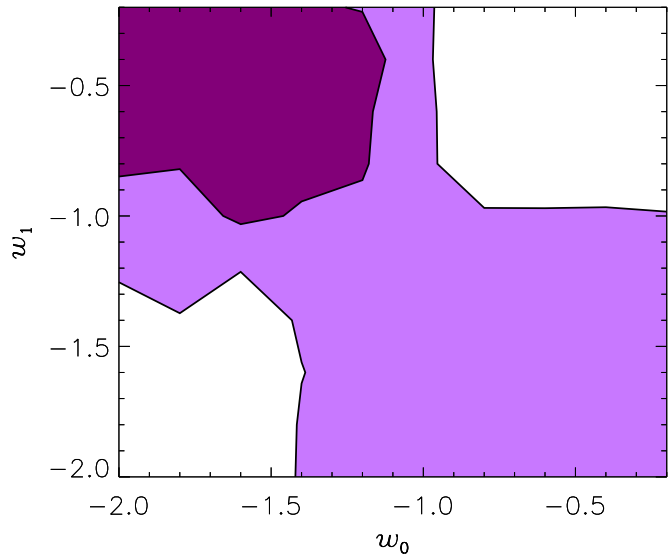}
\caption{68\% (dark) and 95\% (light) likelihood contours for WMAP, SDSS and SNI-a data.}
\label{fig:chi2}
\end{figure}

In Figs.~\ref{fig:chi1}-\ref{fig:chi2} we show results of the likelihood analysis. For each $(w_0,w_1)$ we have marginalized over all other parameters.

From the figure including CMB and LSS data only it is clear that there is an almost complete parameter degeneracy with the same shape as predicted in Fig.~\ref{fig:angle}, i.e.\ it follows the degeneracy inherent in the angular distance integral. With the present data there is no way of discriminating between models with strong time variation of $w$, and models with constant $w$.

When only SN data is used there is also an almost perfect degeneracy which comes from the integral $I$ in Eq.~(\ref{eq:snl}). Again, there is no means of discriminating between a time varying and a constant $w$.

However, once SN, CMB and LSS data are combined this degeneracy is partially broken. From Fig.~\ref{fig:chi2} it is clear that the best fit has $w_1 > w_0$, meaning that $w$ is decreasing with time.
The reason that Fig.~\ref{fig:chi2} is not a superposition of Figs.~\ref{fig:chi1} and \ref{fig:chi3} is that the best fits for the individual subsets of data require different values of $\Omega_m, a_s$ and $q$.
In Table~\ref{table2} we tabulate $\chi^2$ for various best fit models, 
compared with the corresponding best fits for $\Lambda$CDM models $(w_0=w_1=-1)$.
The best fit model to the combined CMB, LSS and SNI-a data has $\chi^2=1623.1$, which should be compared to the 1512 degrees of freedom in the fit. The goodness-of-fit (GoF)\footnote{Defined as GoF = $\Gamma(\nu/2,\chi^2/2)/\Gamma(\nu/2)$, where $\Gamma(\nu/2,\chi^2/2)$ is the incomplete gamma function and $\nu$ is the number of degrees of freedom.} of this model is 0.0236, indicating that only 2.36\% of all randomly generated data sets based on the same underlying cosmological model would produce a worse fit than the one observed. The main cause of this bad fit is the WMAP data, a fact which is well known, and which is presumably due to incomplete foreground removal.

At first sight the models with varying $w$ seem to better fit the data, disfavouring a pure cosmological constant. However, four new parameters are added to the model with time varying $w$. The GoF for the best fit $\Lambda$CDM model is actually 0.0239 ($\chi^2 = 1626.9$ for 1516 degrees of freedom), which is better than for the time-varying $w$ model.
Therefore there is no compelling evidence for a time-variation of $w$ in the present data. We elaborate more on this point in section~\ref{sec:rank} where we perform non-parametric rank correlation tests on the SN data.

\begin{figure}
\hspace*{1.5cm}\includegraphics[width=100mm]{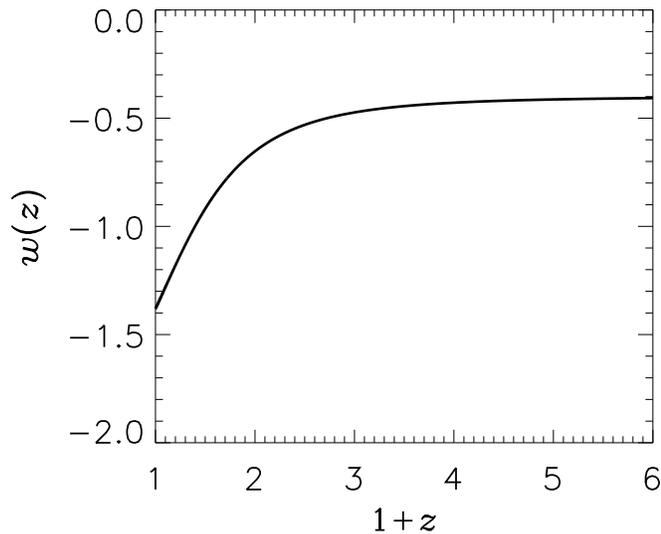}
\caption{Evolution of $w$ in the best-fit model found.}
\label{fig:bestfit}
\end{figure}

Our best fit model has parameters $w_1 = -0.4$, $w_0=-1.8$, $\Omega_m=0.38$, $a_s = 0.50$, and $q=3.41$. In figure~\ref{fig:bestfit} we show $w(z)$ for this model. Evidently the data favour a strongly time-dependent $w$ at present.
We note that this result is similar to that of \cite{Corasaniti:2004}. In this work $w$ is also parameterized as evolving from one asymptotic limit to another, albeit with a different parametrization. Their best fit model is found to have $w_1 = -0.7$, $w_0 = -2.0$, using slightly different observational data than is used in the present work.

\begin{table}[ht]
\caption{\label{table2} $\chi^2$ for the best fit model, compared with $\Lambda$CDM, for different data sets}
\begin{indented}
\item[]
\begin{tabular}{@{}lll}
\br
Data used & best fit & $\Lambda$CDM \cr
\mr
CMB+LSS    &    1446.7    &    1447.3 \\
SNI-a      &     173.8    &    177.1  \\
CMB+LSS+SNI-a &  1623.1   &    1626.9 \\
\br
\end{tabular}
\end{indented}
\end{table}

\subsection{A constant equation of state}

\begin{figure}
\hspace*{1.5cm}\includegraphics[width=100mm]{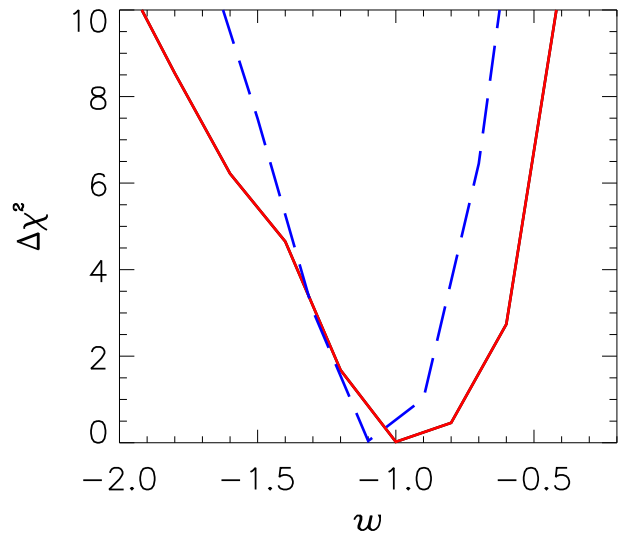}
\caption{Likelihood functions for the case of $w_0=w_1$. The full (red) line is for WMAP+SDSS, and the dashed (blue) for WMAP+SDSS+SNI-a data.}
\label{fig:w1d}
\end{figure}

The simplest possibility for dark energy is that $w$ is constant. Constraints on this type of model can be found by setting $w_0=w_1$, effectively taking line sections through the $(w_0,w_1)$ likelihood plots. The resulting 1D $\chi^2$ curves can be seen in figure~\ref{fig:w1d}. From the combination of all available data we find that $-1.34 \leq w \leq -0.79$, in good agreement with other recent analyses \cite{Knop:2003,Riess:2004}.

\subsection{Parametrization in terms of $w(z=0)$ and $dw/dz(z=0)$}

In fact our parametrization can easily be reduced to a model where the fit is done for $w(z=0) \equiv w_*$ and $dw/dz(z=0) \equiv w_*^\prime$.

\begin{figure}
\hspace*{1.5cm}\includegraphics[width=100mm]{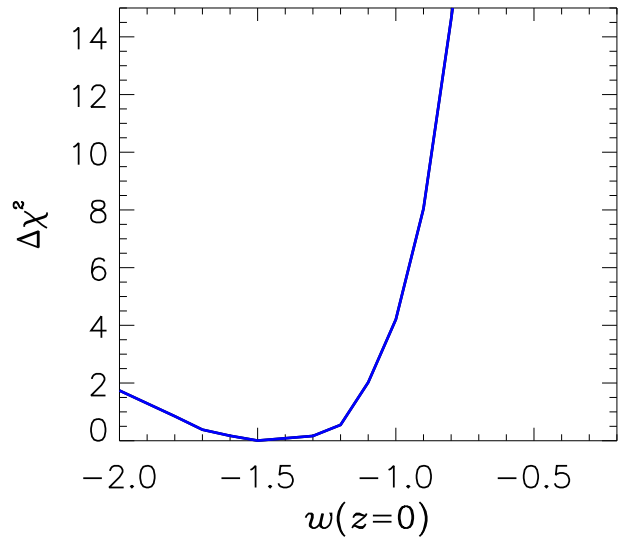}
\caption{$\chi^2$ as a function of the present day value of $w$ for the  WMAP+SDSS+SNI-a data.}
\label{fig:wz0}
\end{figure}

To do a one dimensional fit over $w_*$, one simply rewrites $w_1$ as
\begin{equation}
w_1 = \frac{w_* a_s^q w_0}{w_0 (1+a_s^q) - w_*}
\end{equation}
When the likelihood analysis is done, $w_0$ is a free parameter and $w_1$ given by the relationship above. 
The result of this exercise is shown in Fig.~\ref{fig:wz0}. From this it can be seen that the best fit is at $w_* = -1.43$ as described above, and that $w_* = -1$ is disfavoured at about $2\sigma$, which is also in accordance with our previous finding. The $1\sigma$ allowed interval is
$w_* = -1.43^{+0.16}_{-0.38}$. While the best fit value is very close to what is obtained by Alam et al. \cite{Alam:2004a,Alam:2004b}, and also by Riess et al., the error bars are different from those of Riess et al. The reason is that we add other observational data which shift the best fit value, and also that differences between our parametrization and the Taylor series used in other analyses increase when moving away from the best fit values.
Basically our parametrization has more freedom to fit data because there are additional free parameters. This in turn means that the one-parameter error bars increase.

\begin{figure}
\hspace*{1.5cm}\includegraphics[width=100mm]{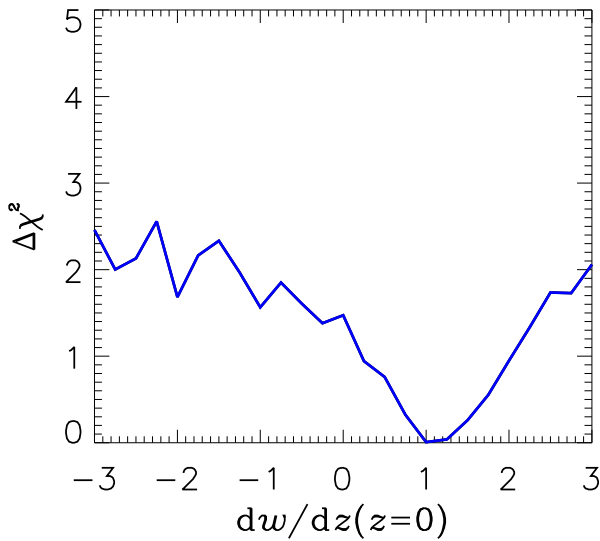}
\caption{$\chi^2$ as a function of the present day value of $dw/dz$ for the combined WMAP+SDSS+SNI-a data.}
\label{fig:derivs}
\end{figure}

Equally, it is possible to do a parametrization in terms of $w_*^\prime$ instead in the following way. First, the derivative of $w$ with respect to $a$ is
\begin{equation}
w_*^\prime = - \left.\frac{dw}{da}\right|_{z=0} =  \frac{w_0 w_1 q a_s^q (w_1-w_0)}{(w_1 + a_s^q w_0)^2}.
\end{equation}
We now introduce a new variable $x=a_s^q$. By doing this, we can solve the above equation for $q$
\begin{equation}
q= w_*^\prime \frac{(w_1 + x w_0)^2}{w_0 w_1 x (w_0-w_1)}.
\end{equation}
In the likelihood analysis, the free parameters are then $w_0,w_1$ and x, whereas $q$ is given by the above relation.
The result of this analysis is shown in Fig.~\ref{fig:derivs}

Formally, the 1$\sigma$ allowed interval is $w_*^\prime = 1.0^{+1.0}_ {-0.8}$. However, as is evident from the figure the likelihood is far from Gaussian, and no useful constraints can be placed at the 2$\sigma$ level.

\subsection{Comparison with analyses based on Taylor series expansions}

Tentative evidence for evolution of $w$ has been found in several
other analyses of supernova data.  Most analyses of SN data use a
Taylor series expansion of $w$ around $z=0$ because no high redshift
data is used. One recent example is the work of Alam et al. \cite{Alam:2004a,Alam:2004b} where the
parametrization
\begin{equation}
\rho_{DE} = A_0 + A_1 (1+z) + A_2 (1+z)^2
\end{equation}
is used.
This can be translated into a relation for $w$
\begin{equation}
w(z)= -1 + \frac{1+z}{3} \frac{A_1 + 2 A_2 (1+z)}{A_0 + A_1 (1+z) +
A_2 (1+z)^2},
\end{equation}
where $A_0 = 1 - \Omega_m - A_1 - A_2$.

The best fit found in that work is $\Omega_m = 0.3$, $A_1=-4.36$, $A_2
= 1.83$ (slightly better fits are found for higher $\Omega_m$, but the
corresponding $A_1$ and $A_2$ are not tabulated).  This leads to
\begin{eqnarray}
w |_{z=0} = -1.33 \,\,\,\,\,\,   &   \,\,\,\,\,\, \left.\frac{dw}{dz}\right|_{z=0} = 1.07
\end{eqnarray}

Our results cannot be compared on global scale because the Taylor
series expansion quickly breaks down (in fact it cannot generally be
expected to be valid even at $z \sim 1$). The only direct comparison
available is in terms of the present $w$ and its derivative.

As described above, we find that the best fit model has
\begin{eqnarray}
w |_{z=0} = -1.4 \,\,\,\,\,\,   &   \,\,\,\,\,\, \left.\frac{dw}{dz}\right|_{z=0} = 1.0
\end{eqnarray}
which is almost identical to what is found by Alam et al.

As discussed previously the reason that $dw/dz \neq 0$ is a better fit
than a constant $w$ is exactly the supernova data, and therefore our
results can be expected to be very similar in the one limit where they
can be directly compared.

\subsection{SNIa data rank correlation test}
\label{sec:rank}

Since the SNIa data are best fit by a changing $w$ and there are
different opinions about the significance of the effect and the
dependence on the specific dark energy model, we have also
investigated the need for additional parameters in the luminosity
distance relation by studying the correlation between the observed
magnitude and the redshift after subtracting the best fit cosmology
magnitudes. This can be done in an non-parametric manner using the
Spearman or Kendall rank correlation tests. The two-sided
significance level, $p$ (ranging from 0 to 1), of the deviation from
zero correlation indicates whether the model is capable of
representing the complexity of the actual data or not. A small value
of $p$ indicates a significant correlation or
anticorrelation\footnote{Note that the zero point magnitude is
irrelevant for both the likelihood analysis and the correlation
tests.}. For an Einstein-de Sitter universe, we get $p\lesssim
10^{-19}$, indicating -- as expected -- that this simple model is
not able to mimic the observed magnitudes over a wide redshift
range. Fitting $\Omega_m$ for a zero cosmological constant universe,
we get $p\sim 0.17-0.19$ (for $\Omega_m=0$), showing a better but
still unsatisfactory fit. Including a cosmological constant with or
without a flat universe assumption removes any evidence of correlation
(see table \ref{tab:corr}) and disfavouring the need for more
complexity in the dark energy model. We have also simulated a sample
100 SNIa data sets with the same number and redshift distribution of
SNe as the Riess gold sample assuming a cosmological constant flat
universe with $\Omega_m = 0.3$ and a dispersion in SN magnitudes of
0.3 mag. Subtracting the best fit cosmological constant flat universe
and performing the rank correlation tests on each simulation yields
\begin{eqnarray}
\Omega_m &=& 0.31\pm 0.05 \nonumber\\
p_S &=& 0.47\pm 0.29 \nonumber\\
p_K &=& 0.47\pm 0.30 .
\end{eqnarray}
Comparing with the results in table \ref{tab:corr} it is clear that the
real data actually have a smaller deviation from zero correlation than
is expected for a cosmological constant universe, thus weakening the
case for a evolving $w$.

This result is completely consistent with the fact that the Goodness-of-Fit for the combined CMB, LSS, and SNI-a analysis is slightly worse for the time-varying $w$ models than for pure $\Lambda$CDM models.

\begin{table}[ht]
\caption{\label{tab:corr} 
Result from likelihood analysis and correlation test. $p_S$ and $p_K$
are the two-sided significance level (ranging from 0 to 1) of the
deviation from zero for the Spearman and Kendall rank correlation
tests. A small value of $p$ indicates a significant correlation or
anticorrelation.}
\begin{indented}
\item[]
\begin{tabular}{@{}llll}
\br
Cosmology & Best fit parameters & $p_S$ & $p_K$\cr
\mr
Einstein-de Sitter & --- &  $10^{-19}$ & 0 \\
Matter dominated & $\Omega_m=0$ & 0.20 & 0.17\\
$\Omega_m +\Omega_\Lambda=1$ & $\Omega_m=0.31$ & 0.88 & 0.86\\
$\Omega_m$ and $\Omega_\Lambda$& $\Omega_m=0.46,\,\Omega_\Lambda = 0.98$ & 0.88 & 0.86\\
Flat with const. $w$ & $\Omega_m=0.49,\,w=-2.4$ & 0.88 & 0.89\\
Flat with linear $w$ & $\Omega_m=0.48,w\,_*=-2.5,w\,_*^\prime =2.0$ & 0.98 & 0.98\\
\br
\end{tabular}
\end{indented}
\end{table}

\subsection{Extending the parametrization} 

The parametrization used in the present analysis already involves more parameters than can be unambiguously fitted with present data. Nevertheless it it still interesting to see how the parametrization naturally extends to a larger number of parameters.

\begin{figure}
\hspace*{1.5cm}\includegraphics[width=100mm]{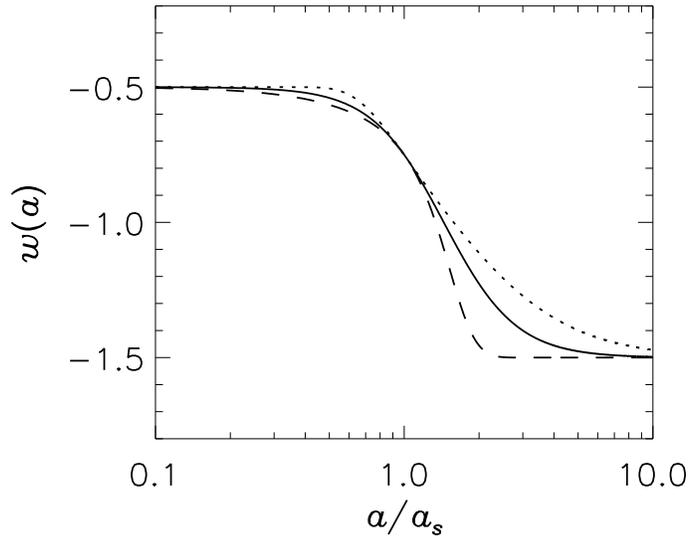}
\caption{$w(a/a_s)$ for different values of $r$. Other parameters have been fixed to $w_0 = -1.5$, $w_1 = -0.5$, $q=3$. The full line is for $r=0$, the dashed for $r=2$ and the dotted for $r=-2$.}
\label{fig:skew}
\end{figure}

If we still use the assumption that the transition is between two asymptotic values of $w$, the next natural parameter to add is a skewness in the transition profile. This can for instance be incorporated with the parametrization
\begin{equation}
w(a) = w_0 w_1 \frac{a^{q +(a/a_s)^r-1} + a_s^q}{a^{q +(a/a_s)^r-1} w_1 + a_s^q w_0},
\end{equation}
where $r$ is a new parameter. For negative $r$ the transition is more rapid at early times, whereas for positive $r$ it is more rapid at late times. In fig.~\ref{fig:skew} we show the effect of adding $r$ as a new parameter.

\begin{figure}
\hspace*{1.5cm}\includegraphics[width=100mm]{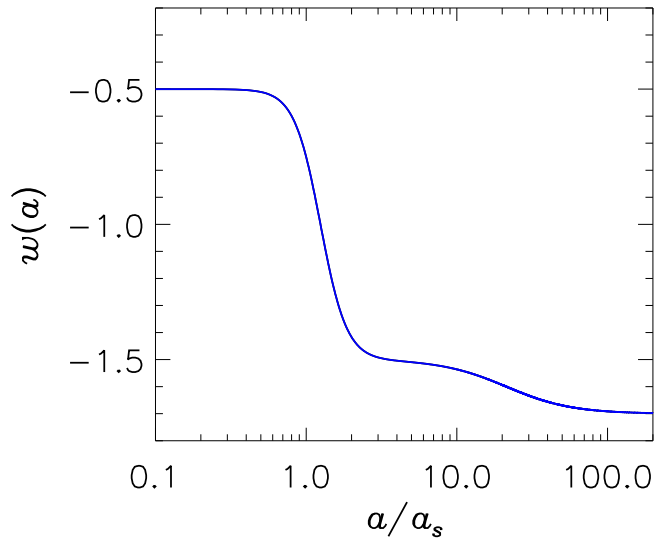}
\caption{$w(a/a_s)$ for a model with two transitions.}
\label{fig:steps}
\end{figure}

Another possibility is that there are several transitions in $w$ during the evolution of the universe. 
This is also simple to incorporate. The initial transition is described by
\begin{equation}
w_i(a) = w_0 w_1 \frac{a^q + a_s^q}{a^q w_1 + a_s^q w_0}.
\end{equation}
To add a second transition, simply take
\begin{equation}
w(a) = w_i(a) w_2 \frac{a^r + a_{s2}^r}{a^r w_2 + a_{s2}^r w_i(a)},
\end{equation}
where $a_{s2}$ is the scale factor at the transition $w_0 \to w_1$, and $r$ describes how rapid the transition is. In figure~\ref{fig:steps} we show a model which has the second transition at $a_{s2} = 20 a_s$, $q=5$, $r=2$, $w_1 = -0.5$, $w_0 = -1.5$, and $w_2 = -1.7$.
By this iterative procedure, more and more steps can be added to the parametrization.

Finally, we note that our parametrization is well-behaved only if $w<0$, even though $w$ can be arbitrarily close to 0. If there is a transition from $w>0$ to $w<0$ there is a singularity in $w$. However, this can easily be cured by taking instead $w(a) = (a^q w_0 + a_s^q w_1)/(a^q+a_s^q)$.

\section{Conclusion} 

We have analysed present cosmological data, using an extension of the current minimal cosmological model to allow for time-variation in the dark energy equation of state, $w$.

The result of the likelihood analysis is that the data favour a $w$
which is currently undergoing a rapid transition towards a more negative value. In our parametrization where $w$ varies between two asymptotic values, the best fit is $w(z \to \infty) = -0.4$, and $w (1+z \to 0) = -1.8$. This result agrees well with other analyses using different parametrizations.

The 1$\sigma$ allowed interval for the present value of $w$ is
$w(z=0)=-1.43^{+0.16}_{-0.38}$ and for its derivative $dw/dz(z=0) = 1.0^{+1.0}_{-0.8}$.

Even though pure $\Lambda$CDM models have higher $\chi^2$ values, they 
also have more fitting parameters. When the Goodness-of-Fit is calculated for both types of models, the best fit $\Lambda$CDM model has GoF = 0.0239, whereas the time-varying $w$ model has GoF = 0.0236. This indicates that there is no real evidence for a time variation of $w$ in the present data.
This finding is corroborated by rank correlation tests on the SNI-a data. From Spearman and Kendalls tests we again find that time-varying $w$ models do not produce a statistically significant improvement in the fit to data.

Finally we have discussed in some detail how to extend our parametrization of $w$ to allow for a more refined description once better data becomes available.

\section*{Acknowledgments} 

We acknowledge use of the publicly available CMBFAST
package~\cite{CMBFAST} and of computing resources at DCSC (Danish
Center for Scientific Computing). SH wishes to thank CERN for support and
hospitality

\vspace*{2cm}

\section*{References} 

\end{document}